\documentclass[letterpaper,titlepage,11pt]{article}
\usepackage{hyperref}
\usepackage{amssymb,amsmath,amsfonts}
\usepackage{epsfig}

\begin{document}

\begin{titlepage}

\rightline{\vbox{\small\hbox{\tt hep-th/0701070} }} \vskip 3cm

\centerline{\Large \bf Entropy of Three-Charge Black Holes on a Circle}

\vskip 1.6cm \centerline{\bf Troels Harmark$\,^{1}$, Kristjan R.\
Kristjansson$\,^{2}$, Niels A. Obers$\,^{1}$, Peter B.
R{\o}nne$\,^{1}$} \vskip 0.5cm
\begin{center}
\sl $^1$ The Niels Bohr Institute  \\
\sl  Blegdamsvej 17, 2100 Copenhagen \O , Denmark \\
\vskip 0.5cm
\sl $^2$ Nordita \\
\sl  Blegdamsvej 17, 2100 Copenhagen \O , Denmark \\
\end{center}

\vskip 0.5cm

\centerline{\small\tt harmark@nbi.dk, kristk@nordita.dk,
obers@nbi.dk, roenne@nbi.dk}

\vskip 1.2cm \noindent
We study phases of
five-dimensional three-charge black holes with a circle in their transverse space.
In particular, when the black hole is localized on the circle we compute the corrections
to the metric and corresponding thermodynamics in the limit of small mass. When taking
the near-extremal limit, this gives the corrections to the finite entropy of
the extremal three-charge black hole
as a function of the energy above extremality. For the partial extremal limit with
two charges sent to infinity and one finite we show that the first correction to the
 entropy is in agreement with the microscopic entropy by taking into account
that the number of branes shift as a consequence of the interactions across the transverse circle.

\end{titlepage}

\section{Introduction}

Three-charge black holes in five dimensions play a prominent role in
string theory. In particular, these black holes are the first
examples where  a microscopic description
\cite{Strominger:1996sh,Callan:1996dv}
of the entropy was achieved in terms of degeneracies of D-brane states.
These three-charge black holes, and the closely related two-charge systems, continue
to provide a fertile ground for further exploration of the microscopic origin of
entropy in string theory (see e.g. the review \cite{Mathur:2005ai}) and, more generally,
the AdS${}_3$/CFT${}_2$ correspondence \cite{Maldacena:1997re}.

In this talk, we review recent results  \cite{Harmark:2006df} on
these three-charge systems in the setting where one of the spatial
dimensions is compactified on a circle, i.e.\ when the
five-dimensional spacetime is asymptotically $\mathcal{M}^4\times
S^1$. Of special interest here is the case when the black hole is
localized on the circle. Considered on the covering space of the
circle, this corresponds to an infinite array of black holes on a
line, separated by a distance equal to the circumference of the
circle. At extremality, it follows from the BPS property that
electric repulsion cancels the gravitation attraction, so that the
entropy is given by the finite result $S=2\pi\sqrt{N_1 N_4
N_0}$. However, when we move away from extremality the black holes
will start to interact with each other and as a consequence the
entropy will receive corrections.  Here we summarize how, in the
large radius limit, the first two leading corrections of this
entropy can be computed by relating five-dimensional three-charge
black holes  on  a circle to neutral Kaluza-Klein black holes (see
the reviews \cite{Harmark:2005pp,Kol:2004ww,Harmark:2007md})
 in five dimensions.
This map also enables us to numerically obtain the entire phase of
three-charge black holes localized on a circle. A more general consequence
of the map is the appearance of a rich phase structure for three-charge black holes on a
circle, including a new phase where the black
holes are non-uniformly smeared on the circle. We also
review the non-trivial fact that, when
two charges are extremal and one is kept finite, the first correction to the macroscopic entropy obtained for the localized case  can be understood
from a microstate counting picture.

\section{Generating three-charge black holes on a circle}

The method we employ to generate the three-charge black holes we are interested in,
builds on the technique used in \cite{Harmark:2004ws}, where it was shown that any
Kaluza-Klein black hole in $d+1$ dimensions ($4 \leq d \leq 9$) can  be mapped to a
corresponding brane solution of Type IIA/IIB String Theory and
M-theory, following the procedure originally conceived in
\cite{Hassan:1992mq}. These are thermal excitations of
extremal 1/2-BPS branes in String/M-theory with
transverse space $R^{d-1} \times S^1$. This gives a precise connection between the rich phase
structure of Kaluza-Klein black holes and that of the corresponding brane.
In particular, by considering the near-extremal limit of the latter,
the thermodynamic behavior of the non-gravitational theories dual to
near-extremal branes on a circle is obtained via the map.

By considering the particular case of Kaluza-Klein black holes in five dimensions
$(d=4)$, the above map can be generalized to generate non-extremal
three-charge brane configurations in Type IIA/IIB String Theory
and M-theory from any neutral five-dimensional Kaluza-Klein black hole.
We generally work in the duality frame where these charges are
carried by the F1-D0-D4 system%
\footnote{Our results take the simplest form
in this duality frame, but we note that by a T-duality in the direction of the F1-string
this is related to the P-D1-D5 system.},
and the solutions obtained are
thermal excitations of the corresponding extremal 1/8-BPS brane
system. When reduced on the spatial world-volume directions of
these branes, we then obtain three-charge black holes in
five-dimensional supergravity.

There is considerable knowledge on the phase structure of five-dimensional
Kaluza-Klein black holes. While there are four known phases, we will
confine ourselves to the following three:%
\footnote{There is also a phase of bubble-black hole sequences (see e.g. \cite{Elvang:2004iz}),
consisting of black holes held apart by Kaluza-Klein bubbles,
and the corresponding three-charge black hole phases will be considered in
\cite{Harmark:2007}.}
 i) The uniform phase, i.e.\ the black
string which is a four-dimensional Schwarzschild black hole times a
circle, so the horizon topology is $S^2 \times S^1$;
ii) The non-uniform phase, which is a static
solution emerging from the uniform phase at the Gregory-Laflamme point
where the black string is marginally unstable.
For five dimensions, the leading order behavior
of this phase was first found numerically in \cite{Gubser:2001ac} and
recently extended into the non-linear regime in Ref.~\cite{Kleihaus:2006ee}.
This phase also has horizon topology $S^2 \times S^1$ but the horizon is non-uniform
along the direction of the circle;
iii) The localized phase, which approaches the five-dimensional
Schwarzschild black hole in the limit of zero mass. The first
correction to this metric in the small-mass limit is known analytically
\cite{Harmark:2003yz} (see also
\cite{Gorbonos:2005px}) using the ansatz proposed in
\cite{Harmark:2002tr}  and recently also the second correction
\cite{Karasik:2004ds,Chu:2006ce}, while the most recent numerical data on the
entire phase can be found in Ref.~\cite{Kudoh:2004hs}.
The horizon topology in this phase is $S^3$.

The three-charge solutions that we are interested in can then be generated by
 using these neutral five-dimensional Kaluza-Klein black holes as seeding solutions.
Starting with any such five-dimensional solution we add five flat
dimensions and then act on the solution with a series of three
boosts and various U-dualities, where each boost adds one charge to the
system. The new solution has metric  \cite{Harmark:2006df}
\begin{equation}
\label{eq:einsteinmetric}
ds_{10}^2 = H_1^{-\frac{3}{4}}H_4^{-\frac{3}{8}}H_0^{-\frac{7}{8}}
    \left( -Udt^2  +H_4H_0 dx^2 + H_1H_0 \sum_{i=1}^4 (du^i)^2 +
             H_1H_4H_0\frac{L^2}{(2\pi)^2} V_{\alpha \beta }dx^\alpha dx^\beta  \right) \ ,
\end{equation}
where $U$ and $(L/2\pi)^2 V_{\alpha \beta }$ ($1 \leq \alpha, \beta \leq 4$) are the metric components
of the five-dimensional seeding solution.
The dilaton and gauge potentials are given by
\begin{align}
\label{eq:dilaton-KalbRamond}
 e^{2\phi} = H_1^{-1}H_4^{-\frac{1}{2}}H_0^{\frac{3}{2}}, \quad
A_{a} = \coth \alpha_a (H_a^{-1} - 1) dt \wedge d\omega_{a},
\textrm{ for } a=1,4,0 \ ,
\end{align}
where $H_a$ are harmonic functions on the transverse space, given by
\begin{equation}
\label{eq:Ha}
H_a = 1+ (1-U) \sinh^2 \alpha_a, \quad \textrm{for }a = 1,4,0 \ .
\end{equation}
This three-charge solution describes a non-extremal configuration
with F1-string, D4-brane and D0-brane charges, with the label $a=1,4,0$ referring to the
type of object. To compare with the five-dimensional black hole, we can compactify again
 the five added dimensions.

Each of the three phases of Kaluza-Klein black
holes described above thus directly maps onto a corresponding phase
of non-extremal three-charge black holes and we can express the
thermodynamic properties in terms of those of the seeding solution.
For the non-extremal three-charge black holes described above,
the physical quantities that we can measure asymptotically are the mass,
$\bar M$, the three charges $Q_a$, and the tension ${\mathcal T}$ in
the transverse direction.
We define dimensionless mass $\bar \mu = g \bar M$ and charge
$q_a = g Q_a$ using the dimensionfull parameter $g$,
defined in Eq.~\eqref{eq:nearextremallimit}, that involves the
Newton's constant, the size of the circle and the volume of the
five new additional dimensions that we take to be compactified on
a torus.
The relative tension is defined as
$\bar n = L{\mathcal T}/(\bar M- M^\textrm{el})$ \cite{Harmark:2004ch} where in the
denominator we subtract the electric part of the mass from the total mass.

By comparing the asymptotic behavior of the seeding solution to the
three-charge solution, we can get a map from $\mu$ and $n$ of the original
neutral black hole \cite{Harmark:2003dg,Kol:2003if} and the boost parameters
$\alpha_a$, to the physical quantities of the new charged solution,
$\bar \mu$, $q_a$ and $\bar n$.
Given the value of the three charges we can write the  mass and relative tension
of the charged solutions as
\cite{Harmark:2006df}
\begin{align}
\bar \mu = \sum_a q_a + \frac{1}{2}\mu n
    + \frac{(2-n)\mu }{6}\sum_a \frac{b_a}{1+\sqrt{1+b_a^2}},
 \quad
b_a \equiv \frac{2-n}{6}\frac{\mu}{q_a} \;\;\;,\qquad \bar n = n \ .
\end{align}
The neutral seeding solution
always has $\mu \ge 0$ and  $0 \le n \le 2$ so the last two terms are
positive, and therefore we see that for fixed $q_a$,
the mass $\bar \mu$ is bounded from below by the sum of the charges, which
is the extremal mass.
The energy above extremality will thus  be defined as the mass minus the sum of the charges.

\section{Near-extremal limit and thermodynamics}

The map above becomes especially simple (as was the case for
the one-charge branes considered in Ref.~\cite{Harmark:2004ws})
when we take the near-extremal limit. This limit is also relevant
for the dual CFT description of these brane systems and for our
application to microstate counting. The near-extremal limit we take has
the property that the energy above extremality has the same scale as the
size of the circle, and is defined as
\begin{align}
\label{eq:nearextremallimit}
L \to 0, \quad
\alpha_a \to \infty, \quad
\ell_a \equiv L^{\gamma_a} \sqrt{q_a}  \textrm{ fixed}, \quad
g \equiv \frac{16 \pi G_{10}}{V_1V_4 L^2}  \textrm{ fixed} \ .
\end{align}
Here $\gamma_a$ are positive constants that sum to one. When one of them is taken to
be zero the corresponding charge remains finite in the near-extremal limit.
In order to get finite results, we must rescale all background fields with appropriate powers
of $L$  and in order for this rescaling to be a symmetry of the
action, the 10-dimensional Newton's constant needs to be rescaled so that $g$
remains fixed.

Taking the limit (\ref{eq:nearextremallimit}) on the non-extremal three-charge solution
(\ref{eq:einsteinmetric}), (\ref{eq:dilaton-KalbRamond}) produces a new solution of
near-extremal three-charge black holes on a circle, in which essentially the constant
``1'' in the harmonic functions (\ref{eq:Ha}) has disappeared (see \cite{Harmark:2006df}
for details). The asymptotic physical quantities for the near-extremal solution
are the energy above extremality $\epsilon = gE$, which is the mass minus the
charges, and the tension of the circle $\hat{\mathcal T}$.
The relative tension is now defined as
$r = 2\pi \hat{\mathcal T}/E$.
The map from $(\mu,n)$ of the neutral seeding solution to $(\epsilon,r)$ of
the near-extremal three-charge solution
turns out to be surprisingly simple \cite{Harmark:2006df}
\begin{align}
\label{eq:themap}
\epsilon = \mu n/2, \quad r = 2 \ .
\end{align}
It is interesting to notice that the relative tension is independent
of the parameters of the original black hole. This means that the three
phases all collapse on top of each other in the $(\epsilon,r)$ diagram.
This result is intimately related to the finite non-vanishing entropy in the
extremal limit \cite{Harmark:2006df}. Furthermore,
given the rescaled entropy ${\mathfrak s}$ and temperature ${\mathfrak t}$ of the seeding solution,
we can find the rescaled entropy $\hat {\mathfrak s}$ and temperature $\hat {\mathfrak t}$
of the near-extremal solution via the map  \cite{Harmark:2006df}
\begin{align}
\hat {\mathfrak t} = {\mathfrak t} ({\mathfrak t}{\mathfrak s})^{3/2},
\quad \hat {\mathfrak s} = {\mathfrak s} ({\mathfrak t}{\mathfrak s})^{-3/2} \ .
\end{align}
Reinstating the units for the rescaled entropy we obtain the relation
$S= \hat {\mathfrak s} \  2\pi\sqrt{N_1N_4N_0}$ where $N_a$, $a=1,4,0$
are the number of each type of extended object in the system.

\begin{figure}
\includegraphics[width=0.4\columnwidth]{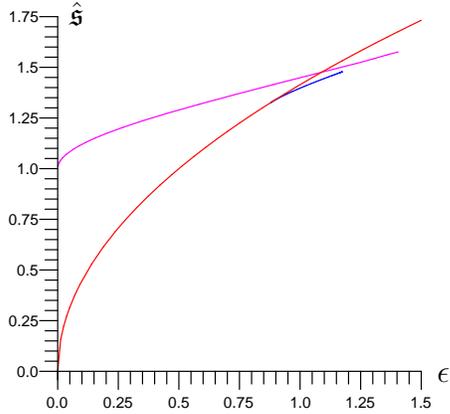}
\put(-1,25){\large $\epsilon$}
\put(-140,160){\large$\hat{\mathfrak s}$}
\caption{The entropy $\hat {\mathfrak s} = S
/(2\pi\sqrt{N_1N_4N_0})$ as a function of the energy above
extremality $\epsilon$ for the localized phase (magenta), the
uniform phase (red) and the non-uniform phase (blue).  The curves
for the localized/non-uniform phase are based on the numerical
data of \cite{Kudoh:2004hs} and \cite{Kleihaus:2006ee}
respectively. We see that the localized phase starts out at finite
entropy. It has higher entropy than the uniform phase and is
therefore thermodynamically stable. At some point the curves of
the localized and the uniform phases meet and the non-uniform
phase interpolates between them.} \label{fig:svse}
\end{figure}

We can now apply the map to the known thermodynamics of each of the
three phases for the neutral seeding solution, and obtain the
thermodynamics of the corresponding three phases of
three-charge black holes. As an example, Fig.~\ref{fig:svse} shows
the entropy as a function of energy in the near-extremal case for each
of these three different phases, using numerical data of
Ref.~\cite{Kudoh:2004hs,Kleihaus:2006ee}.
Moreover, from analytic results for small localized Kaluza-Klein black holes
\cite{Harmark:2003yz,Gorbonos:2005px,Karasik:2004ds,Chu:2006ce}
we get an expansion for the entropy of localized near-extremal three-charge
black holes on a circle  \cite{Harmark:2006df}
\begin{align}
S= 2\pi\sqrt{N_1N_4N_0}
    \left(1+\sqrt{\frac{\epsilon}{8}}+\frac{\epsilon}{16}
     + {\mathcal O}(\epsilon^{3/2}) \right)\ .
\end{align}
This central result presents the first two corrections to the finite
extremal entropy \cite{Strominger:1996sh}, in a small energy (or equivalently,
large circle radius) expansion.

We can also consider other near-extremal limits, for example
keeping one of the three charges finite while sending the other two
to infinity, by taking one of the $\gamma_a$ in (\ref{eq:nearextremallimit}) to be zero.
Choosing $\gamma_0=0$, corresponding to keeping finite D0-brane charge, we then find for the localized black hole the energy above extremality
 and entropy  \cite{Harmark:2006df}
\begin{align}\label{eq:thermoonefinite}
\epsilon & =  \rho_0^2\sinh^2\alpha_0
    + \frac{1}{2} \rho_0^2 \left( 1 + \frac{1}{16} \rho_0^2\right)
    + {\cal{O}} (\rho_0^6)\ , \\
\hat {\mathfrak s} &= \rho_0 \cosh \alpha_0
    \left(1 + \frac{1}{16} \rho_0^2 + \frac{1}{512} \rho_0^4 \right)
    + {\mathcal O}(\rho_0^7) \ .
\end{align}
Here $\rho_0$ is a small parameter of the original seeding solution, related
to the location of the event horizon and thus controlling
the size of the neutral black hole relative to the circle radius.  In the next section
we will review how the expression for the entropy can be reproduced from
a microstate counting point of view.

\section{Microscopic counting of entropy}

We now want to recalculate the entropy using microstate
counting techniques.  First, we recall how the microstate entropy
formula is generalized to non-extremal black holes that are not on
a circle.
In the weak coupling limit, the extremal black hole can be
described as a configuration with $N_4$ D4-branes, a number $N_1$
of fundamental strings stretching between the D4-branes and $N_0$
D0-branes that are threaded on the fundamental strings
\cite{Strominger:1996sh}.
Non-extremal black holes can be generated by, for example, adding a
number $N_{\bar 0}$ of anti-D0-branes.
Horowitz, Maldacena and Strominger \cite{Horowitz:1996ay}
argued that in the dilute gas
limit, the forces between the D0- and anti-D0-branes would be very
small and interactions could be ignored.  The entropy is thus given
by the sum of the square-roots \cite{Horowitz:1996ay}
\begin{align}
\label{eq:sumofsqrt}
S= 2\pi\sqrt{N_1N_4}\left(\sqrt{N_0} + \sqrt{N_{\bar 0}}\right)\ .
\end{align}
This is the formula that we want to generalize to our three-charge
black holes on a circle.
In the non-extremal formula the D0- and anti-D0-branes were
far apart and did not interact, but with the small transverse circle
present, that is no longer a safe assumption to make.
In our near-extremal limit (\ref{eq:nearextremallimit}), the size of the transverse circle is taken
to be at the same scale as the energy above extremality and the
interaction energy is therefore not negligible compared to the excitation
energy.  That means that interactions of the zero-branes across the
circle must be taken into account.
The effect of the interaction is to shift the number of D0-branes for
a given total energy \cite{Costa:2000kf}.
To find the entropy we must therefore find the effective number of
branes $N_0'$ and $N_{\bar 0}'$.

To this end we write the total mass of our three-charge system as
\begin{align}
\label{eq:deltaEVint} \bar M = Q_1 + Q_4 + \delta E + V_\textrm{int} \ ,
\end{align}
where $\delta E$ is the energy carried by the D0-branes and the
anti-D0-branes, and $V_\textrm{int}$ is the interaction energy
related to the presence of the transverse circle.  As we start
adding anti-D0-branes to the extremal system, they will interact
with the D0-branes across the circle, reducing their energy by
$V_\textrm{int} = {\mathcal T} L/2$.
We find the effective number of D0- and anti-D0-branes from
requiring
\begin{align}
\label{eq:deltaENbarN}
\delta E  \simeq N'_0 + N'_{\bar 0},
\quad Q_0 \simeq  N'_0 - N'_{\bar 0} \ .
\end{align}
Applying this method to the localized phase of the three-charge
black hole on a circle and inserting the effective numbers into
Eq.~\eqref{eq:sumofsqrt} we obtain the entropy
\begin{align}
S = \frac{2\pi \sqrt{N_1 N_4N_0}}{\ell_0}  \rho_0 \cosh\alpha_0
        \left(1+ \frac{\rho_0^2}{16}+\mathcal{O}(\rho_0^4)\right).
\end{align}
We hence find that the leading order correction of this
microscopic entropy is in agreement with the leading order correction of the macroscopic entropy
in (\ref{eq:thermoonefinite}).

\section{Conclusion}

We have shown how to generate five-dimensional three-charge
black hole solutions on a circle, from known phases of five-dimensional
Kaluza-Klein black holes and considered the near-extremal
limit of these. In particular, we have computed the first corrections to the finite entropy
of (extremal) localized three-charge black holes and matched, in a
partial near-extremal limit, the macroscopic entropy with a
microscopic calculation. We believe this is a non-trivial check on the
microscopic picture employed here, while at the same time illustrating
the power of the map. We have also obtained an entirely new phase of
non- and near-extremal three-charges black holes, which are non-uniformly
distributed along the circle.

One open direction is to further examine the microscopic description of these
black holes. In this connection, the entropy matching discussed here
was recently extended to second order  in Ref.~\cite{Chowdhury:2006qn}.
Furthermore, in Ref.~\cite{Chowdhury:2006qn}
a simple microscopic model was proposed that reproduces most of the features
of the phase diagram, including the new non-uniform phase. The study of other
new three-charge solutions, arising from bubble-black hole sequences will also
be interesting and we intend to report on this in the future \cite{Harmark:2007}.
Finally, further examination of the non-uniform phase of three-charge
black holes and the classical stability of the uniform phase along the lines of
\cite{Harmark:2005jk}, and its relation to correlated stability conjecture would be
interesting to pursue.

\section*{Acknowledgement}
NO would like to thank the organizers of the RTN workshop in Napoli, October 9-13, 2006
for the opportunity to present this work.
 Work partially supported by the European Community's Human Potential
Programme under contract MRTN-CT-2004-005104 `Constituents,
fundamental forces and symmetries of the universe'.




\providecommand{\href}[2]{#2}\begingroup\raggedright\endgroup
\end{document}